\documentclass[pdflatex,sn-mathphys-num]{sn-jnl}

\usepackage{graphicx}%
\usepackage{multirow}%
\usepackage{amsmath,amssymb,amsfonts}%
\usepackage{amsthm}%
\usepackage{mathrsfs}%
\usepackage[title]{appendix}%
\usepackage{xcolor}%
\usepackage{textcomp}%
\usepackage{manyfoot}%
\usepackage{booktabs}%
\usepackage{algorithm}%
\usepackage{algorithmicx}%
\usepackage{algpseudocode}%
\usepackage{listings}%

\theoremstyle{thmstyleone}%
\theoremstyle{thmstyletwo}%

\theoremstyle{thmstylethree}%

\begin{document}

\title[Article Title]{Programmable spatiotemporal OAM optical toroidal beams with completely tunable properties} 


\author*[1]{\fnm{Andrew V.} \sur{Komonen}}\email{a.komonen@uq.edu.au}
\author[2]{\fnm{Nicolas K.} \sur{Fontaine}}
\author[1]{\fnm{Martin} \sur{Pl\"oschner}}
\author[1]{\fnm{Marcos Maestre} \sur{Morote}}
\author[2]{\fnm{David T.} \sur{Neilson}}
\author[1]{\fnm{Joel} \sur{Carpenter}}
\author*[1]{\fnm{Mickael} \sur{Mounaix}}\email{m.mounaix@uq.edu.au}


\affil[1]{\orgdiv{School of Electrical Engineering and Computer Science, The University of Queensland, Brisbane, QLD, 4072, Australia}}

\affil[2]{\orgdiv{Nokia Bell Labs, 600 Mountain Avenue, Murray Hill, NJ 07974, USA}}


\abstract{Spatiotemporal toroidal orbital angular momentum (OAM) beams are a developing class of spatiotemporal beams which have key applications within quantum physics, metrology, imaging and optical manipulation. However, the full realization of these applications require complete configurability within tunable temporal duration, 3D geometric structure and OAM charge of these beams along with amplitude, phase and polarization control. In this paper, we demonstrate complete configurability of programmable, polarization-resolved OAM toroidal beams after propagation through a multimode optical fiber (MMF) supporting 90 spatial/polarization modes. We show high fidelity control: temporally with beams spanning 2.3 ps - 6.8 ps, geometrically with toroidal aspect ratios spanning 1.5-2.7 and with up to $|l|=13$ OAM topological charge. In total this system supports 25,000 spatiotemporal and polarization degrees of freedom which enables the independent control of all physical and geometric properties of these 3D toroidal beams. By utilizing an MMF, this system also enables toroidal beam delivery to previously inaccessible regions, paving the way for applications including  optical manipulations, sensing and imaging through complex photonics media such as scattering biological tissues.}

\maketitle

\section{Introduction}\label{sec1}

\begin{figure} [H]
  \centering
  \includegraphics[width=\textwidth]{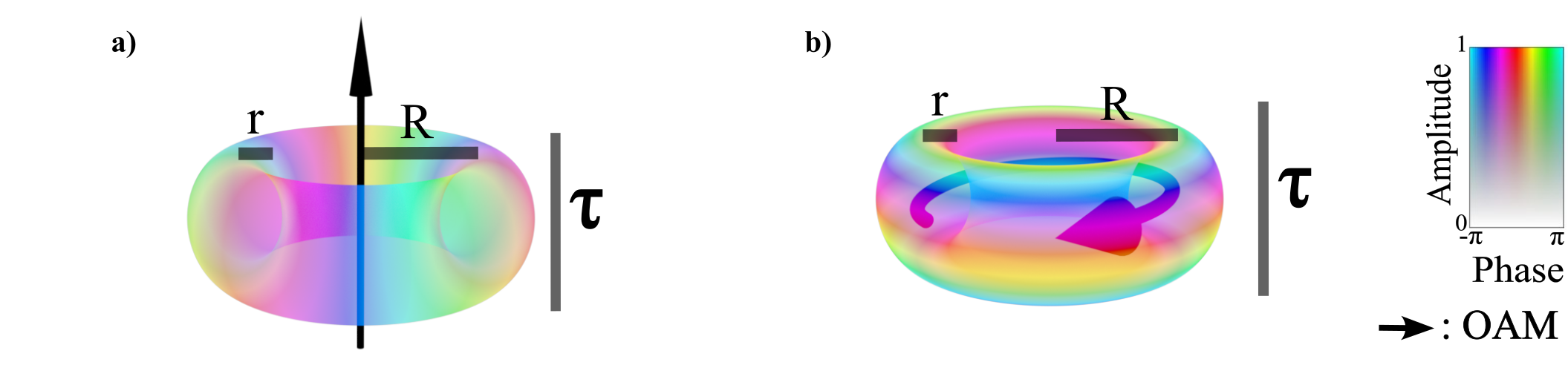}
  \caption{Conceptual optical toroidal beam and their tunable properties. a) Toroidal phase wrap with topological charge $l$=+3. The OAM is passing through the toroidal singularity b) Poloidal phase wrap with topological charge $l$=+3. The OAM is internally circulating. $r$ and $R$ represent respecitvely the minor and major radii of the torus, governing its aspect ratio $\frac{R}{r}$. $\tau$: temporal duration of the toroidal beam}
  \label{fig:torus}
\end{figure}

Structured light beams which have been sculpted in space and time are known as spatiotemporal light beams. These beams are expanding applications across photonics~\cite{roadmap_spatiotemp_light} including for optical communications~\cite{wang_optical_comms}, nonlinear optics~\cite{jolly_propagation_2025, vieira_nonlinear_2014} and metrology~\cite{cheng_metrology_2025}. Toroidal beams form a subset of spatiotemporal light beams. These beams have a donut shaped intensity in 3D where this donut geometry is completely defined in space and time, as illustrated in Fig.~\ref{fig:torus}. These beams also have the capability to contain orbital angular momentum (OAM) through a spirally varying phase encoding around the toroidal surface~\cite{wan_review, zhan_spatiotemporal_tutorial}. Two common encodings include around the poloidal direction, that is around the toroidal tube, which generates internally circulating OAM~\cite{allen_orbital_1992,speirits_waves_2013,wan_toroidal_2022} and the toroidal direction, that is around the center hole of the toroid, which generates OAM passing through this center hole. 



Toroidal beams have recently been realized in the acoustic domain~\cite{liu_generation_2025}, but their generation in optics remain significantly more challenging than producing 2D defined space-time optical vortices (STOVs)~\cite{jolly_propagation_2025, jhajj_spatiotemporal_2016_first, hancock_free-space_2019, piccardo_broadband_2023, chong_generation_2020_transverse, adams_spatiotemporal_2024}. Only recently the first demonstrations of free-space optical toroidal beams have been reported~\cite{wan_toroidal_2022, zdagkas_observation_2022}. In our previous work, we reported the generation of toroidal beams in arbitrary 3D orientations in space, time and polarization, delivered after propagation through a multimode optical fiber (MMF)~\cite{komonen_spatiotemporal_2025}, by enabling full vectorial spatiotemporal control~\cite{mounaix_time_2020}. However, many emerging applications, such as for optical manipulation, imaging and quantum communication, demand not only spatiotemporal and polarization control, but also complete configurability of both the physical and geometrical properties of these beams~\cite{huang_integrated_2025, shen_optical_2019, lei_optical_2024}. For instance, in high-dimensional quantum entanglement, information can be encoded in the OAM degree of freedom~\cite{huang_integrated_2025, shen_optical_2019, erhard_advances_2020}. Combining OAM with other degrees of freedom, such as their temporal duration and polarization, enables access to higher-dimensional quantum states~\cite{barreiro_generation_2005}. This enables the implementation of advanced quantum communication protocols and logic quantum gates, including the deterministic CNOT gate~\cite{zhu_implementation_2025}. 

A key remaining degree of freedom for full beam configurability is the 3D geometric structure of the toroidal beam itself. Specifically, the structure is defined by its major radius ($R$, distance from toroid center to tube center) and minor radius ($r$, radius of the tube)~\cite{moroni_toric_2017_toroid_maths}. Combined these two radii define the toroidal aspect ratio as $\frac{R}{r}$ which directly define the beam shape. Control over the 3D geometric structure enables enhanced performance in applications such as optical manipulation ~\cite{liu_monolithic_2025} and light-matter interactions~\cite{begin_orbital_2025}. When combined with polarization control, these beams also enable depth-resolved 3D imaging with improved performance~\cite{tan_minflux_2025}, and enhances sensing applications, such as for turbulent media, where different beam geometries interact differently with local refractive index fluctuations~\cite{chen_weather_2025}. 

In addition to geometry, control over OAM and temporal duration further expands the range of toroidal beam applications. Tuning the OAM charge enables enhanced light-matter interactions such as for the absorption, dispersion~\cite{kudriasov_azimuthally_2025} and ionization of matter~\cite{begin_orbital_2025}, key mechanisms for quantum sensing applications~\cite{kudriasov_azimuthally_2025}. Higher-order OAM modes improve resolution and transmittance through scattering media such as biological tissue~\cite{tan_minflux_2025} while also enhancing sensitivity to turbulent flows~\cite{chen_weather_2025}. Temporal tuning might also provide benefits for optical manipulation and trapping of nanoscale particles~\cite{chiang_femtosecond_2013, shane_effect_2010,kittiravechote_enhanced_2014}. A fully tunable toroidal beam could offer a new platform to investigate optical trapping dynamics.

In this paper, we introduce a fully configurable platform for generating 3D toroidal beams, with independent control over their temporal duration, OAM charge and geometric structure. Such parameters are introduced in Fig.~\ref{fig:torus}. Our experimental apparatus (see Methods), also enables control over the beam's amplitude, phase and polarization, after propagation through a graded-index multimode fiber (MMF) which supports 90 spatial/polarization modes~\cite{komonen_spatiotemporal_2025}. Having control over all these modes enables us to decouple and tailor both OAM and beam geometry, which is not possible with single Laguerre-Gaussian (LG) modes alone. This level of control constitutes a complete spatiotemporal beam-shaping toolbox, enabling new frontiers in high-dimensional quantum entanglement, quantum sensing, metrology, imaging and optical manipulation.

\section{Results}\label{sec2}

\subsection{Configuring Time Duration}

\begin{figure} [H]
  \centering
  \includegraphics[width=0.95\textwidth]{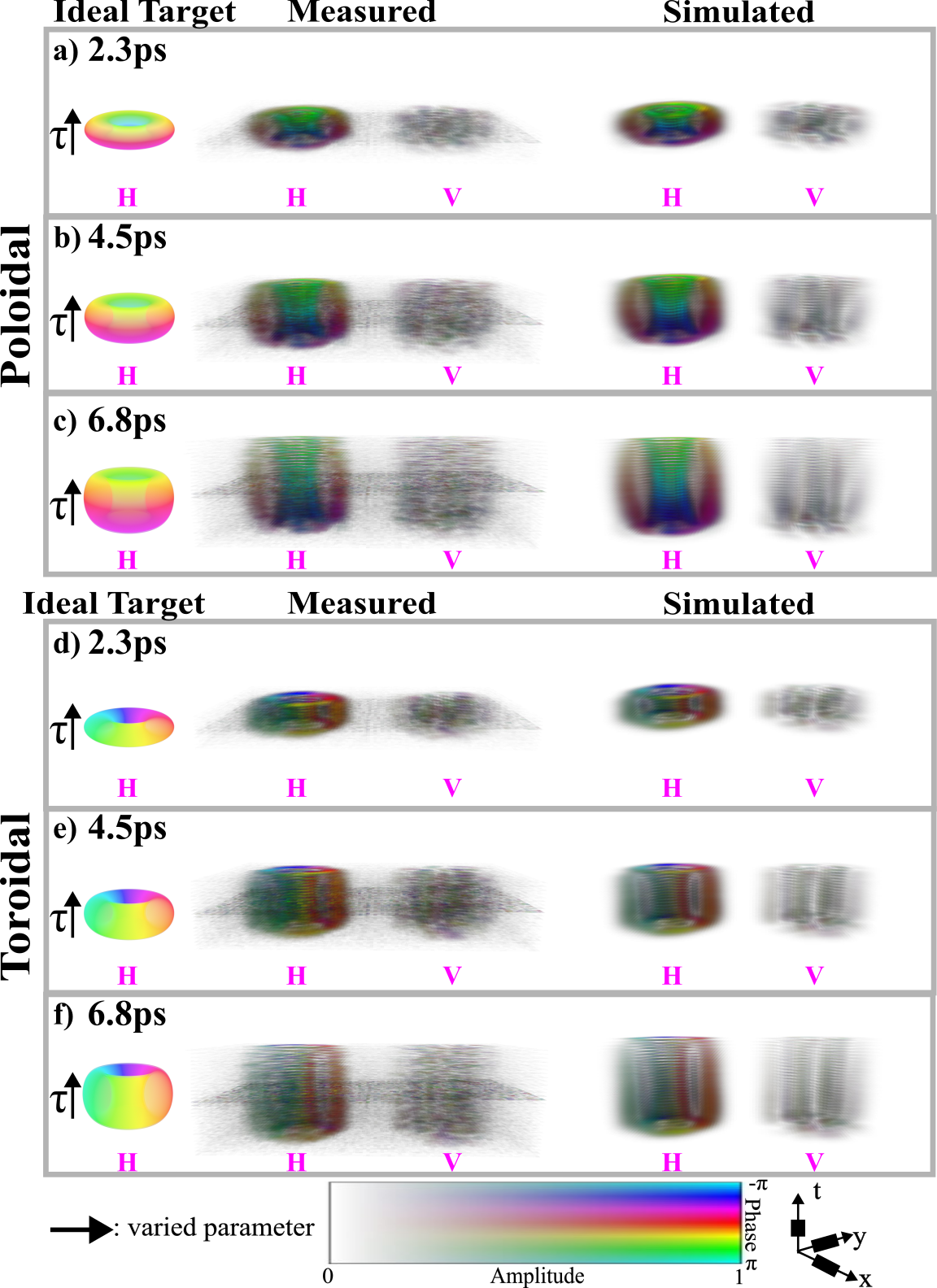}
  \caption{Temporal duration control of toroidal beams with different phase wraps. a-c) toroidal beams with a poloidal phase wrap. d-f) toroidal beams with a toroidal phase wrap. The temporal duration for each beam is 2.3 ps, 4.5 ps and 6.8 ps respectively. All beams have a phase wrapping of $2\pi$, corresponding to a topological charge of 1, and an aspect ratio of 2. Axis ticks: $x$ and $y$: 500$\mu$m, $t$: 1ps}
  \label{fig:Time Duration}
\end{figure}

3D toroidal beams with different temporal durations are shown in Fig.~\ref{fig:Time Duration}. These beams are all generated in the horizontal polarization. Fig.~\ref{fig:Time Duration}a)-c) are 2.3 ps, 4.5 ps and 6.8 ps in duration respectively and contain poloidal phase wrapping of topological charge +1 which generates internally circulating OAM. Fig.~\ref{fig:Time Duration}d)-f) have the same time durations as Fig.~\ref{fig:Time Duration}a)-c) but they show a different phase encoding of the toroidal beam, as they contain toroidal phase wrapping of topological charge +1. This toroidal phase wrapping generates OAM that passes through the beam's singularity.

To evaluate the quality of these beams, we have included numerically propagated beams which appear under "Simulated" in Fig.~\ref{fig:Time Duration}. For each configuration, these simulated beams were generated by first constructing the targeted toroidal beam as a superposition of 45 Hermite-Gaussian modes, and then numerically propagating it using the experimentally measured TM~\cite{komonen_spatiotemporal_2025}. Therefore, this provides the best possible toroidal beam the experimental system can achieve. We then compared the experimentally measured beams ("Measured" in Fig.~\ref{fig:Time Duration}) with the corresponding simulated beam using an overlap integral ($\mathcal{O}$)~\cite{komonen_spatiotemporal_2025}, which quantifies similarity in amplitude, phase and polarization between the "Measured" and "Simulated" beams. The magnitude-squared of the overlap integral $|\mathcal{O}|^2$ serves as a quality metric. For the poloidal phase wraps in Fig.~\ref{fig:Time Duration}a)-c) we measured overlaps of $|\mathcal{O}|^2 = 86\%$, $|\mathcal{O}|^2 = 78\%$ and $|\mathcal{O}|^2 = 66\%$ respectively. For the toroidal phase wraps in Fig~\ref{fig:Time Duration}d)-f) we measured overlaps of $|\mathcal{O}|^2 = 79\%$, $|\mathcal{O}|^2 = 69\%$ and $|\mathcal{O}|^2 = 55\%$ respectively. Interestingly, the beam quality for both phase wraps seem to decrease as the temporal duration increases. This occurs due to the longer temporal durations corresponding to narrower spectral bandwidth. Which means that as the bandwidth narrows, the SLM hologram must define increasingly complex beam structures with fewer spectral components, effectively using fewer SLM pixels. This leads to greater sensitivity to pixel crosstalk, and results in reduced beam quality~\cite{lazarev_beyond_2019,mounaix_time_2020}.

We also observe, for the same time durations, that toroidal beams with toroidal phase wraps exhibit lower overlap values than with poloidal phase wraps. This difference arises from the increased complexity involved in encoding toroidal phase wrapping with the SLM hologram. This is due to the toroidal phase wraps requiring a continuously varying transverse phase across each temporal cross-section, which unavoidably increases the hologram complexity, amplifying the impact of pixel crosstalk, and thus degrading the beam quality.  In contrast, for the poloidal phase wrap, each temporal cross-section has a uniform phase, which corresponds to a simpler SLM hologram with smaller pixel crosstalk effects and hence greater beam quality. Overall, despite these experimental challenges, these results highlight the system's capabilities to generate high-quality toroidal beams with time durations of up to 6.8 ps.

\subsection{Configuring Geometric Structure}

\begin{figure} [H]
  \centering
  \includegraphics[width=\textwidth]{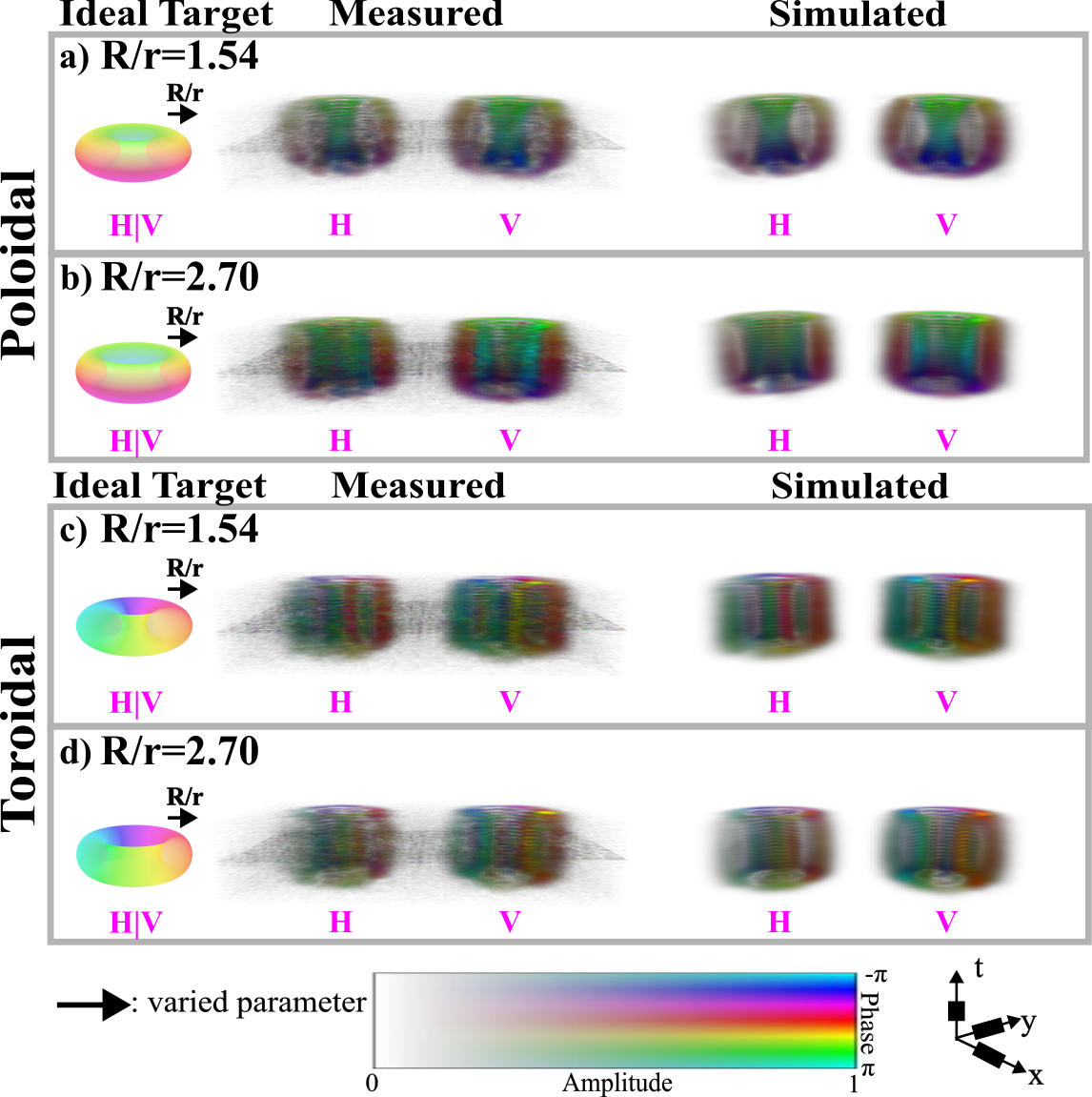}
  \caption{Geometric structure control of toroidal beams. Toroidal beams with varying aspect ratios, where $\frac{R}{r}=1.54$ for a) and c), and $\frac{R}{r}=2.70$ for b) and d), which changes the geometrical structure of the central hole. a-b): Beams with poloidal phase wrapping. c-d) Beams with toroidal phase wrapping. All beams have a temporal duration of 4.5 ps, and a total phase wrapping of $2\pi$, which corresponds to a topological charge of 1. Axis ticks: $x$ and $y$: 500$\mu$m, $t$: 1 ps}
  \label{fig:Radius}
\end{figure}

Toroidal beams with configurable geometric structure are shown in Fig.~\ref{fig:Radius}, with the beams generated for both horizontal and vertical output polarizations simultaneously. The geometric structure is controlled by varying the aspect ratio $\frac{R}{r}$. 
Fig.~\ref{fig:Radius}a)-b) show toroidal beams with aspect ratios $\frac{R}{r}=1.54$ and $\frac{R}{r}=2.70$ respectively, both encoded with a poloidal phase wrapping of topological charge +1, which generates internally circulating OAM. Fig.~\ref{fig:Radius}c)-d) show the same respective minor radii, but the beams have a toroidal phase wrapping of topological charge +1, resulting in an OAM that passes through the beam singularity. All beams presented in Fig.~\ref{fig:Radius} have a temporal duration of 4.5 ps. The toroidal beams with poloidal phase wraps in Fig.~\ref{fig:Radius} have overlaps of $|\mathcal{O}|^2 = 78\%$ and $|\mathcal{O}|^2 = 76\%$ respectively, while the toroidal beams with toroidal phase wraps yield overlaps of $|\mathcal{O}|^2 = 71\%$ and $|\mathcal{O}|^2 = 70\%$ respectively. As observed previously in Fig.~\ref{fig:Time Duration}, the toroidal beam with toroidal phase wraps exhibit slightly lower overlaps due to the increased complexity of the SLM holograms.

Such results are interesting, especially considering that the aspect ratios were chosen to test the system's limits (See Supp. Sec. 3). For instance, with $\frac{R}{r}=1.54$, the hole of the toroid becomes significantly compressed. as shown in Fig~\ref{fig:Radius}a) and c). This value approaches the upper limit of the system's capability, using 90 polarization/spatial modes. Any further increase of $r$ results in light filling the hole of the toroid, hence  affecting the structure of the beam. Conversely, $\frac{R}{r}=2.70$ approaches the lower limit of the system. As shown in Fig.~\ref{fig:Radius}b) and d), the hole of the toroid expands such that the inner and the outer rings of the beam's structure start to converge. Beyond this point, the beam transitions from a toroidal structure to a cylindrical geometry, compromising the target beam structure (See Supp. Sec. 3). Despite the system's inherent limitations, Fig.~\ref{fig:Radius} confirms that the system can robustly generate a wide range of toroidal geometries with high beam qualities, supporting aspect ratios ranging from $1.54 - 2.74$. The achievable aspect ratio is not limited by the choice of phase wrapping on the surface of the toroidal beam, but rather by the number of spatial/polarization modes that define the beam in the transverse direction. Furthermore, the ability to simultaneously generate such toroidal beams for both horizontal and vertical polarizations while maintaining overlaps greater than $70\%$ demonstrates the system's polarization control.

\subsection{Configuring OAM topological charge}

\begin{figure} [H]
  \centering
  \includegraphics[width=\textwidth]{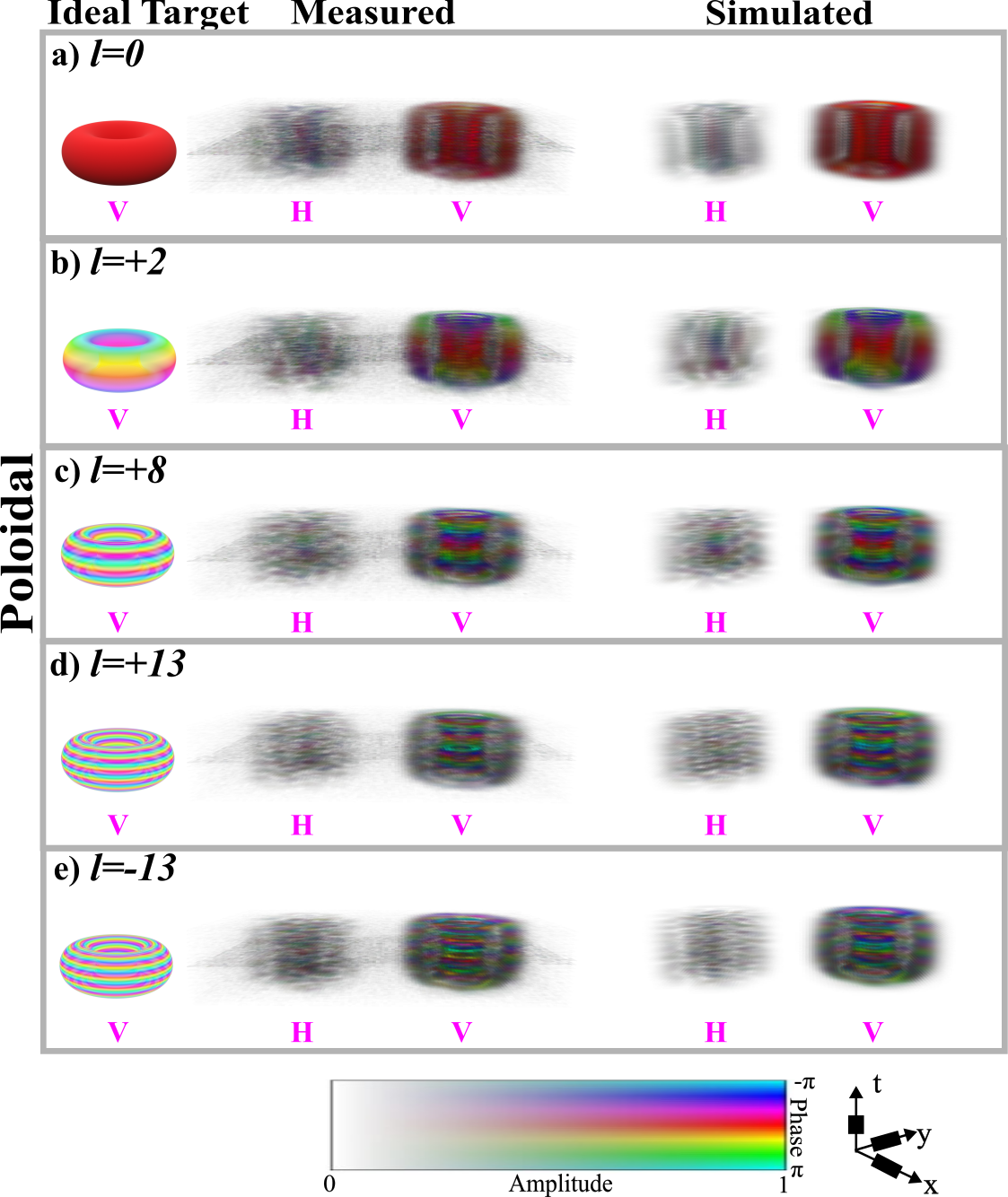}
  \caption{Toroidal beams with varying OAM topological charges with poloidal phase wrapping. The OAM topological charge is set to 0,+2,+8,+13 and -13 respectively. Hence, the corresponding total phase wrapping is $0$, $+4\pi$, $+16\pi$, $+26\pi$ and $-26\pi$ respectively. All of these beams are 4.5 ps in duration and have a minor and an aspect ratio of 2. Axis ticks: $x$ and $y$: 500$\mu$m, $t$: 1ps}
  \label{fig:OAM poloidal}
\end{figure}

\begin{figure} [H]
  \centering
  \includegraphics[width=\textwidth]{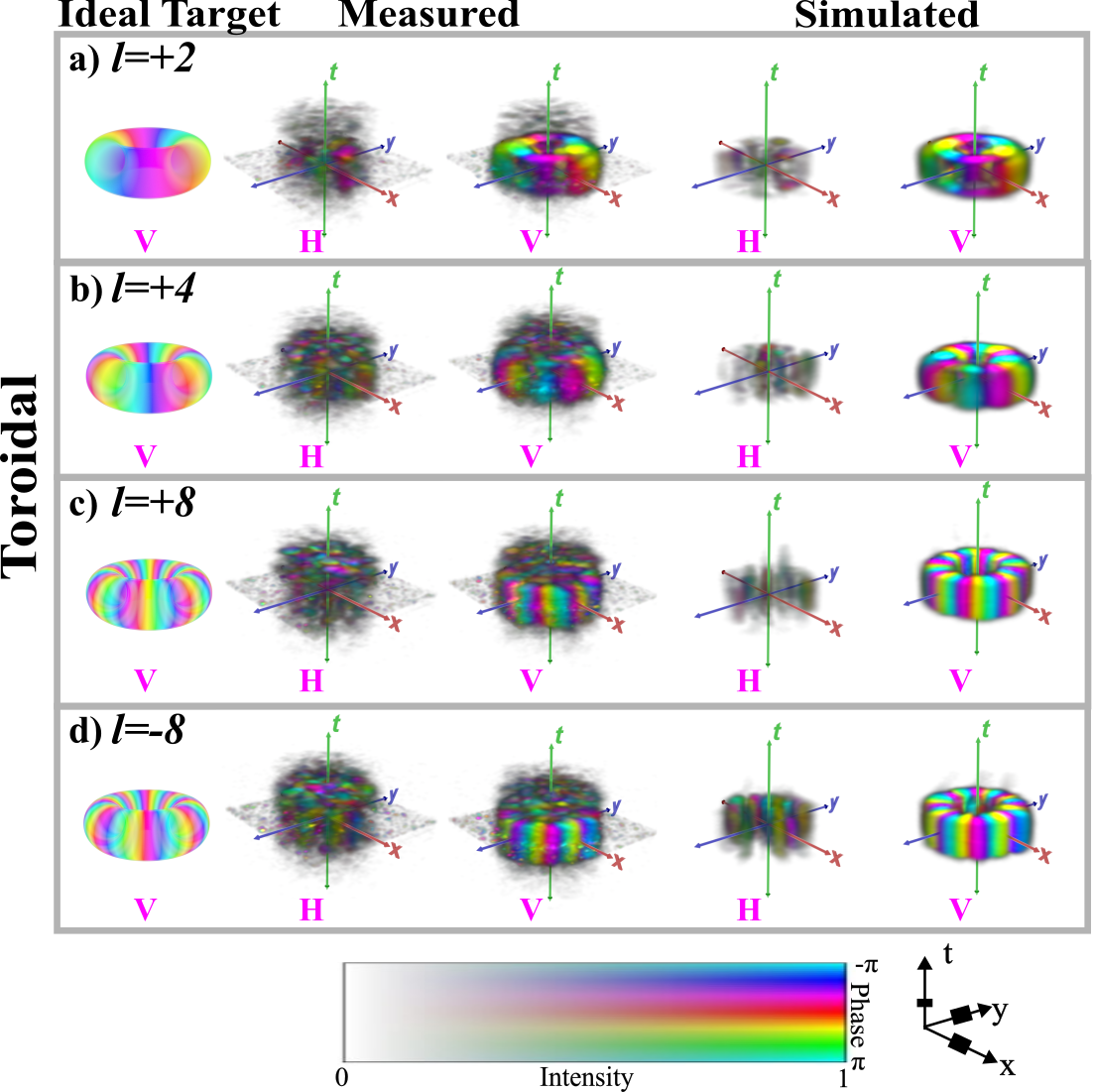}
  \caption{Toroidal beams with varying OAM topological charges with toroidal phase wrapping. The OAM topological charge is set to +2,+4,+8 and -8 respectively. Hence, the total phase wrapping is $+4\pi$, $+8\pi$, $+16\pi$ and $-16\pi$ respectively. All of these beams are 4.5 ps in duration and have an aspect ratio of 2. Axis ticks: $x$ and $y$: 500$\mu$m, $t$: 1ps}
  \label{fig:OAM toroidal}
\end{figure}

Toroidal beams with configurable OAM topological charge are shown in Fig.~\ref{fig:OAM poloidal} and Fig.~\ref{fig:OAM toroidal} for poloidal and toroidal phase wraps respectively, where all beams are generated in the output vertical polarization. 

In Fig.~\ref{fig:OAM poloidal}a-e), the topological charge is set to 0, +2, +8, +13 and -13 respectively. All beams have a poloidal phase wrap, which generates internally circulating OAM. Each beam has a temporal duration of 4.5 ps, and an aspect ratio of 2. The beam overlaps are $|\mathcal{O}|^2 = 79\%$, $|\mathcal{O}|^2 = 83\%$, $|\mathcal{O}|^2=81\%$, $|\mathcal{O}|^2=75\%$ and $|\mathcal{O}|^2=65\%$ respectively. These results demonstrate the system's capability to generate a wide range of OAM values, up to 27 distinct topological charges from $l=$-13 to +13 with high beam quality sustained up to the system's limit of $\pm$13. The limiting factor to further increasing the OAM lies in the number of discrete temporal cross-sections used to define the beam. In this case, we use 20 discrete time steps, which define 20 phase points along the outer and 20 along the inner edge of the toroidal tube, giving a total of 40 points to define the phase wrap. As the topological charge ($l$) increases, the total phase change around the toroidal tube increases by $l(2\pi)$~\cite{wan_toroidal_2022}. For $l=\pm13$, we have just over 3 steps per full $2\pi$ phase wrap, which is just over the minimum required phase sampling to ensure the desired OAM is generated~\cite{vijayakumar_generation_2019}. For large topological charge, the beam quality drops due to the increased phase gradient between adjacent temporal cross-sections, which require more complex holograms to accurately generate the target beam. Such holograms make the system more sensitive to pixel crosstalk on the SLM. 

In Fig.~\ref{fig:OAM toroidal}a-d), the topological charge is set to + 2,+ 4,+ 8 and -8 respectively and these toroidal beams have a toroidal phase wrap. These beams have the same duration, major radius and minor radius as the poloidal OAM beams presented in Fig.\ref{fig:OAM poloidal}. The beam overlaps are $|\mathcal{O}|^2 = 76\%$, $|\mathcal{O}|^2=69\%$, $|\mathcal{O}|^2=69\%$ and $|\mathcal{O}|^2=67\%$ respectively. This confirms that the system is able to generate 17 OAM topological charges on the toroidal phase, spanning up to $l=\pm$8, with good beam quality. The primary limitation with increasing the toroidal OAM charge beyond $|l|=8$ is the Hermite-Gaussian basis of 45 modes per polarization that defines the transverse beam. Such basis cannot generate a topological charge  $|l|>9$ on the toroidal axis, simply because the system would need higher order Hermite Gaussian modes to create it (see Supp. Sec. 1).

\section{Discussion}

In this paper, we have shown full customization of both geometric and physical properties of 3D toroidal beams. This includes precise control over time duration, geometric structure (via the toroidal aspect ratio) and OAM for both poloidal and toroidal phase wraps. We demonstrated the maximum limits of our system, successfully generating toroidal beams with durations ranging from 2.3 ps to 6.8 ps, aspect ratios $\frac{R}{r}$ ranging from 1.5 to 2.7, poloidal OAM charges up to $\pm$13 and toroidal OAM charges up to $\pm$8. This complete control enables precise beam tailoring for varying applications such as high-dimensional quantum entanglement~\cite{huang_integrated_2025, shen_optical_2019, erhard_advances_2020}, optical imaging~\cite{tan_minflux_2025,shen_optical_2019} and optical manipulation~\cite{chiang_femtosecond_2013, usman_optical_2012, kittiravechote_enhanced_2014}.

The level of control demonstrated here could also be extended to create time-varying OAM beams which have been previously demonstrated for microwaves~\cite{zhang_generation_2023}, extreme ultraviolet (EUV) light~\cite{rego_generation_2019} and near infrared (NIR) light on the femtosecond timescale~\cite{de_oliveira_subcycle_2025}. Our system could enable time-varying OAM beams on the picosecond timescale which could be applicable for nanoscale optical manipulation and trapping~\cite{chiang_femtosecond_2013, usman_optical_2012, kittiravechote_enhanced_2014}.


The ability to rapidly and independently fully configure beam duration, geometric structure and OAM charge addresses the increasingly complex demands of spatiotemporal beam shaping. From a practical perspective, any property of the beam can be reprogrammed on demand via the SLM, without the need for physical reconfiguration of the apparatus. Additionally, by integrating an MMF, the system allows delivery of these beams into hard-to-access environments such as biological tissues. Together with the control over arbitrary orientations of the toroidal beam \cite{komonen_spatiotemporal_2025}, this work establishes a complete and versatile beam-shaping toolbox, ready for deployment in advanced photonic systems through complex, scattering and turbulent environments.

\section{Methods}
\subsection{Experimental setup}
The setup utilizes a swept laser with 4.4 THz of bandwidth (35 nm) centered at 193 THz (1551.6 nm) with a spectral resolution of 15 GHz, providing $N_\lambda$=293 controllable spectral channels. This corresponds with a constant experimental temporal resolution of 226~fs, meaning that each 2D temporal beam cross-section is spaced 226~fs apart. Each 15 GHz segment is addressed on a liquid crystal on silicon (LCoS) spatial light modulator (SLM), contained within a multi-port spectral pulse shaper~\cite{komonen_spatiotemporal_2025, neilson_wavelength_2006}. It also beam shapes each spectral component into a 1D 45 spot array defined in amplitude and phase where each spot is subsequently mapped into an orthogonal Hermite-Gaussian (HG) mode through a multi plane light converter (MPLC)~\cite{fontaine_laguerre-gaussian_2019}. Therefore, through this process, the system supports the generation of any combination of 45 orthogonal Hermite-Gaussian (HG) modes defined in amplitude and phase for each supported wavelength. Spectral phase control through the WSS provides the system with temporal control. Polarization control is enabled by addressing each orthogonal polarization on respective halves of the SLM. The 3D toroidal beam is defined in 2D temporal cross-sections, which are accessible indirectly through the provided spectral control. Each cross-section is constructed from a superposition of 45 orthogonal HG modes, individually defined in amplitude, phase and polarization, collectively forming the complete 3D beam structure~\cite{mounaix_time_2020}.

The setup also utilizes a multimode fiber (MMF) to deliver the toroidal beams to the detection section of the device. The intensity of the field is measured with an infrared (IR) camera. To retrieve the phase, the toroidal beam interferes with a tilted reference beam. Processing the resulting interferogram captured on the camera can retrieve the amplitude and phase distribution of the toroidal beam's field through polarization and spectrally resolved off-axis digital holography~\cite{carpenter_digholo_2022, mounaix_control_2019_TM}. 
The whole system is calibrated by measuring its spectrally resolved linear transmission matrix (TM)~\cite{popoff_measuring_2010_TM}.  For each wavelength, this matrix maps each input mode to the corresponding output field as measured on the camera. The TM is measured sequentially by first displaying a hologram on the SLM to excite a single input mode of the system. Then sweeping the laser in wavelength to record the spectrally-resolved field on the camera. The measured output field is decomposed on a 45-mode Hermite-Gaussian basis for each polarization, which gives $N_{\text{output}}=90$ output modes. Hence each input mode is uniquely transformed through the system by a $(N_\lambda,N_{\text{output}})$ matrix into a unique spectrally-resolved output field. The process is iterated for all the $N_{\text{input}}=90$ spatial/polarization input modes, providing the TM of size $(N_\lambda,N_{\text{output}},N_{\text{input}})$~\cite{mounaix_time_2020}. The measured TM includes the modal dispersion and mode coupling induced by the MMF. Therefore, utilizing the conjugate transpose of this TM~\cite{mounaix_spatiotemporal_2016} allows us to calculate the required input field into the MMF which generates the desired toroidal beams at the output of the MMF~\cite{komonen_spatiotemporal_2025}.

\subsection{Plotting the figures}

In Fig.~\ref{fig:Time Duration}, Fig.~\ref{fig:Radius} and Fig.~\ref{fig:OAM poloidal}, the amplitude of each beam is plotted with linear transparency, with the phase encoded in the RGB color. The 3D beams are plotted as a sequence of temporal cross sections, which helps to see the internal part of the toroidal with poloidal phase wrapping.

In Fig.~\ref{fig:OAM toroidal}, we plotted the data using a volumetric plot instead, as there's more importance in highlighting the toroidal phase wrapping on the structure of the beam rather than visualizing the internal part of the beam. To improve the clarity, we applied a quadratic transparency profile and excluded the points corresponding to the lower 7\% of power across both polarizations.

\backmatter

\section*{Declarations}

\subsection*{Data availability}
The measured data are available from the corresponding authors upon reasonable request.

\subsection*{Acknowledgements}
The authors acknowledge Daniel Dahl for designing and producing the 3D printed mount for the SLM and helpful discussions. The authors also acknowledge Jorge Silva for helpful discussions. We acknowledge the Discovery (DP170101400, DE180100009,
DE210100934, FT220100103, FT230100388) program of the Australian Research
Council, the Westpac Scholars Trust, and NVIDIA Corporation for the donation of
the GPU used for this research.

\subsection*{Author contributions}
Experiments performed by A.V.K. with assistance from M.M.M. Optical system designed by N.F., D.N., J.C. and M.M. Spectral pulse shaper section designed by D.N., N.F. and M.M, MPLC by N.F. and J.C., spatiotemporal holograms designed by A.V.K., M.M., N.F., J.C. and M.P., Optical system assembled by M.M, A.V.K., and N.F and aligned by A.V.K. Data analysis by A.V.K, M.M., N.F., M.P. and J.C. Manuscript written by A.V.K. and M.M. with input from all authors. M.M. conceived and supervised the project.

\subsection*{Ethics declarations}
The authors declare no competing interests.

\subsection*{Supplementary information}
Supplementary Notes 1-4, Figs. 1-4.




\end{document}